\documentstyle[preprint,aps,eqsecnum,prb]{revtex}

\def\Q{\hbox{\lowercase{$q$}}}
\def\qn{q^{-{1\over2}}}
\def\q{q^{1\over2}}
\def\qst{(q^{{1\over2}})^*}
\def\qpm{q^{\pm{1\over2}}}
\def\N{{N\over2}}

\def\r#1{\cite{#1}}

\def\a{{a^\dagger}}
\def\bd{{b^\dagger}}
\def\am{{a_{-}}}
\def\ap{{a_{+}}}
\def\sh{{\sinh}}
\def\ns{{\sin}}

\def\TT{{\theta^{n-1}\over[n-1]_q!}}
\def\sn{{\sum_{m=0}^{n-1}}}
\def\sr{{\sum_{m=0}^{r}}}
\def\snw{{\sum_{m=1}^{n-1}}}
\def\snn{{\sum_{m=1}^{n}}}

\def\pt{{\partial\over\partial\theta}}

\def\dz{\partial_z}
\def\qq{\exp({2\pi i\over n})}
\def\ql{\lim_{q\to\qq}}
\def\qml{\lim_{\q\to-1}}
\def\qpl{\lim_{\q\to1}}
\def\ad{a^\dagger}
\def\ff{f(\theta)}
\def\ed{\epsilon {\cal D}_L}
\def\zz{\vert0\rangle}

\def\gg{{q^{g(A_m)}}}
\def\ssn{{\sum_{n=0}^\infty}}
\def\sq{{\sum_{m=1}^\infty}}
\def\ss{{\sum_{m=0}^\infty}}
\def\pr{\prod_{m=1}^{r}}
\def\pn{\prod_{m=1}^{n-1}}
\def\be{\begin{equation}}
\def\ee{\end{equation}}
\def\eq#1{{(\ref{#1})}}

\def\ie{{\it i.e.}}

\catcode`@=11
\newdimen\z@ \z@=0pt
\def\m@th{\mathsurround=\z@}
\def\ialign{\everycr{}\tabskip\z@skip\halign} 
\def\eqalign#1{\null\,\vcenter{\openup\jot\m@th
  \ialign{\strut\hfil$\displaystyle{##}$&$\displaystyle{{}##}$\hfil
      \crcr#1\crcr}}\,}
\def\matrix#1{\null\,\vcenter{\normalbaselines\m@th
    \ialign{\hfil$##$\hfil&&\quad\hfil$##$\hfil\crcr
      \mathstrut\crcr\noalign{\kern-\baselineskip}
      #1\crcr\mathstrut\crcr\noalign{\kern-\baselineskip}}}\,}
\catcode`@=12

\begin{document}
\begin{flushright}
DAMTP/96-57\,,\,
FTUV/96-39\,,\, IFIC/96-47
\vskip0pt hep-th/9610087
\end{flushright}
\vspace{1cm}
\begin{center}
\begin{Large}
{\bf Geometrical foundations of fractional}\vskip0pt
{\bf supersymmetry}
\\
\end{Large}
\vspace{1cm}
{\bf R.S. Dunne, A.J. Macfarlane}, 
\vskip10pt
{\it Department of Applied Mathematics \& Theoretical Physics}\vskip0pt
{\it University of Cambridge, Cambridge CB3 9EW}
\\
\vskip 10pt
\&
\vskip10pt
{\bf J.A. de Azc\'arraga and J.C. P\'erez Bueno
\footnote{e-mails: r.s.dunne@damtp.cam.ac.uk; a.j.macfarlane@damtp.cam.ac.uk;
\\ 
$\phantom{ca.emails: }$azcarrag@evalvx.ific.uv.es; pbueno@lie.ific.uv.es}}
\\
\vskip10pt
{\it Departamento de F\'{\i}sica Te\'orica and IFIC,}\vskip0pt
{\it Centro Mixto Universidad de Valencia-CSIC}\vskip0pt
{\it E-46100-Burjassot (Valencia) Spain.}
\end{center}
\begin{abstract}
A deformed $q$-calculus is developed on the basis of an algebraic
structure involving graded brackets. A number operator and left and right shift
operators are constructed for this algebra, and the whole structure is related 
to the algebra of a $q$-deformed boson. 
The limit of this
algebra when $q$ is a $n$-th root of unity is also studied in detail. 
By means of a chain rule expansion, the left and right 
derivatives are identified with the charge $Q$ and covariant derivative $D$
encountered in ordinary/fractional supersymmetry and this leads to 
new results for these operators.
A generalized Berezin integral and
fractional superspace measure arise as a natural part of our formalism.
When $q$ is a root of unity the algebra is found to have a non-trivial 
Hopf structure, extending that associated with the anyonic line. 
One-dimensional ordinary/fractional superspace is identified
with the braided line when $q$ is a root of unity, so that 
one-dimensional ordinary/fractional supersymmetry can be viewed as 
invariance under translation along this line. In our construction of 
fractional supersymmetry the $q$-deformed bosons play a role exactly analogous to 
that of the fermions in the familiar supersymmetric case.
\end{abstract}
\vfill\eject

\section{Introduction}

Recently two methods of generalizing ordinary supersymmetry (SUSY) have 
received considerable attention. The generalization to parasupersymmetry
\r{RS,FV,BD,KhareI,KhareII,BDN}  
involves the replacement of the usual bilinear
SUSY algebra with a trilinear algebra in analogy to the way in which the 
ordinary bosonic and fermionic algebras are generalized to those associated 
with parabosons and parafermions \r{Green,OK,GM}. 
Such generalizations involve the 
introduction of a parasuperfield and parasuperspace, and the natural
variables to use in working with these are the paragrassmann variables. 
On the other hand, our main interest in this paper is the
generalization to what is known as fractional supersymmetry (FSUSY) 
\r{BF,ABL,Kerner,Abramov,FIK,ISAEV,DurandI,DurandII,DeberghIII,WSC,Mohammedi,AM,FR} 
which seeks to replace the $Z_2$-grading associated with 
the SUSY algebra  with a $Z_n$-graded algebra in such a way that the FSUSY 
transformations mix elements of all grades. This involves the introduction of 
fractional superfields and fractional superspace.
It is natural when constructing such fields to use a generalization of 
the ordinary Grassmann variables
which is distinct from the paragrassmann variables mentioned above. These 
satisfy $\theta^n=0$ and we follow \r{AM} in referring to them as generalized 
Grassmann variables. To work with such variables it is
necessary to define a generalized Grassmann calculus, including an integral. 
Such a calculus is characterized by a single variable $q$, which is
a root of unity, and one of its central features is that it involves two 
derivatives, one associated with $q$, and the other with $q^{-1}$. Many of 
the problems encountered when trying to
develop and understand the calculus are associated with the fact that $q$ 
is a root of unity. To circumvent these, our treatment begins by generalizing 
to the generic $q$ case. Then by excluding all  $q$  which are roots of 
unity we are able to develop a consistent formalism for what we call 
$q$-calculus. 
In the second half of the paper we extend this formalism to the case of
$q$ a primitive $n$-th root of unity by taking the limit of the generic case. 
The result is a consistently formulated calculus for generalized Grassmann
variables, which naturally includes the ordinary $Z_2$-graded Grassmann 
calculus. 
The structure of this provides us with
new insights into both 
fractional and ordinary supersymmetry.

We begin by establishing in sec. II a grading scheme and graded bracket. 
We then 
introduce the $q$-calculus algebra, comprising a $q$-variable $\theta$ and 
two derivation operators ${\cal D}_L$ and ${\cal D}_R$ with which it can be consistently 
equipped. 
We have normalized these operators in such a way that, when they are
placed within our graded bracket, they induce left and right 
$q$-differentiation. Consistency conditions
between the commutation and grading properties of the coefficients on power 
series expansions of functions of $\theta$ are derived, and in general these 
force us to restrict our attention to either left or right differentiation. 
We introduce left and right $q$-integrals, defined by analogy
with ordinary integrals so that they invert the effect of $q$-differentiation. 
Having established the basic 
structure of $q$-calculus in section III we go on in section IV to
construct shift operators for $q$-variables. In section V we obtain series 
expansions of the number operator, and of functions of this. This enables us 
to establish various algebraic identities including the relationship between 
the left and right derivatives. In section VI we use these 
results to connect the $q$-calculus algebra to the defining algebra of the 
$q$-deformed bosons \r{AC,Macfarlane,Biedenharn}. 
When in section VII we take the limit of 
$q$-calculus in which $q$ is a primitive  $n$-th root of unity 
($n\ne 1$), 
we find that consistency considerations force us to set $\theta^n=0$.
Our approach is valid for each primitive root, though for convenience we use 
$q=$exp$({2\pi i\over n})$ in this paper.

One very important limit
is that of the quantity ${\theta^n\over[n]!}$, which we denote by the symbol 
$z$. It turns out that $z$ behaves just like an ordinary (undeformed, `bosonic') 
variable, and that its derivative also arises naturally (and 
with the properties commonly expected of it) in this limit.
In section VIII we discuss the process by which 
$f(\theta)\to f(z,\theta)$ 
(a fractional superfield) as $q\to\qq$, and give some simple examples. 
We further see that under a left or right shift, $z$, suitably normalized,
transforms exactly like the time variable used in theories of 
ordinary/fractional supersymmetry. Partial derivative expansions of the 
$q\to\qq$ limits of ${\cal D}_L$ and ${\cal D}_R$ show that these are related 
similarly to 
the supercharge/fractional supercharge and covariant derivatives
encountered in such theories, and this implies new results for such operators.
By using $z$ and $\dz$ we are able to give a complete theory of 
$q$-differentiation and $q$-integration in the $q\to\qq$ limit.
By considering the conversion of this algebraic $q$-integral 
into a numerical integral
measure in a way analogous to the undeformed case, we obtain in 
section IX a measure on 
fractional superspace which generalizes the usual
integral measure on superspace, and includes a fractional generalization
of the Berezin \r{Berezin} integral. We are also able to extract 
from our $q$-integral an alternative integral for generalized Grassmann 
variables which has previously appeared in the literature \r{FIK}. 
Thus our point
of view unifies these two approaches.
\par
In sections X and XI
we examine the $q\to\qq$ limits of the shift and number operators, and of 
the relationship of the calculus to the $q$-deformed bosons.
The number operator decomposes into a linear combination of $N_\theta$ and 
$N_z$, where $N_z$ is just the number operator associated with an ordinary 
bosonic degree of freedom. 
The series expansions of the shift operators
also terminate, but they do not decompose like the number operators, into independent $z$ and
$\theta$ parts. 
This is an important point, since because of the interpretation of the coproduct of a braided Hopf 
algebras as a shift operator, it suggests that the braided Hopf structure of fractional supersymmetry
is non-trivial, and motivates a discussion of this.\par
The defining relationship of calculus on
the braided line is in fact identical to the relationship \eq{(11)} between 
$\theta$ and the left derivative ${\cal D}_L$. It is the existence of this 
underlying structure which gives our various constructions their
interesting properties. 
We reproduce the Hopf structure of this line in our own 
notation, and take its limit as $q\to\qq$ to obtain new results. 
The resulting braided Hopf algebra generated by ($z,\theta$), 
which has a richer structure than does the anyonic line \r{MajidII,MAJBOOK}, 
cannot be decomposed into independent $z$ and $\theta$ parts, and therefore 
should not be regarded as a composite entity, but rather as something quite 
new. 
Also, although $z$ and $\dz$ satisfy the algebra associated with ordinary 
calculus, $z$ has a  non-primitive coproduct, a result which also 
follows through for the time variable in ordinary one-dimensional 
supersymmetry. 
It is usual to view the odd supersymmetry transformations as mixing 
the odd and even sectors of a (product) superspace. 
Our results provide
a new geometric interpretation, in which the odd and even sectors of 
 one-dimensional supersymmetry  together make up the braided
line at $q=-1$, and supersymmetry is no more than invariance under translation
along this line. A similar result holds for the fractional case.
In fact, this braided Hopf structure is consistent with the central extension
description of the ordinary supersymmetry group \r{AA} and of fractional
supersymmetry in general \r{AM}.
Higher dimensional supersymmetries and fractional supersymmetries can be
constructed from the same braided point of view, 
 although there are extra subtleties involved\r{RD2}. 
We include appendices in which matrix representations of $q$-calculus,
results to do with our limiting procedure, and identities in $q$-analysis
are derived.
In view of the importance of the limiting process in the work 
described in this paper, we mention that similar procedures have
played a vital role in the development of the theory of quantum groups
at roots of unity. Ideas which stem from the work of 
de Concini, Kac and collaborators, and 
Lusztig appear in two recent monographs: see chapter nine of \r{CP} and chapter 
seven of \r{GRS}, from which they can be traced back to the original references.

\section{Brackets and \Q-grading}

In this section we will establish the bracket notation and grading scheme to 
be used throughout. Let $q,r,s$ and $t$ be arbitrary complex numbers, and 
begin by defining the bracket
\be 
[A,B]_{q^r}:=AB-q^r BA\ =-q^r[B,A]_{q^{-r}}\quad,
\label{(1)}
\ee
and noting the identities
\be
[AB,C]_{q^t}=A[B,C]_{q^{t-r}}+q^{t-r}[A,C]_{q^r} B\quad, 
\label{(2)}
\ee
\be
[A,BC]_{q^t}=[A,B]_{q^{t-r}}C+q^{t-r}B[A,C]_{q^r}\quad,
\label{(3)}
\ee
which are valid for any $t$ and $m$. 
The most general form of the Jacobi identity involving this kind of bracket is
\be
[[A,B]_{q^r},C]_{q^s}+q^{-t}[[B,C]_{q^{r+s+t}},A]_{q^t}+
q^s[[C,A]_{q^{r+t}},B]_{q^{-{(s+t)}}}=0\quad, 
\label{(4)}
\ee
valid for arbitrary numbers $r,s,t$. 
If we assign an integer grading $g(X)$ to each element 
$X$ of some algebra, such that $g(1)$=0 and
\be 
g(XY)=g(X)+g(Y)\quad, 
\label{(5)}
\ee
for any $X$ and $Y$, then we can define a graded bracket as follows, 
\be 
[A,B]_\chi:=AB-q^{-g(A)g(B)}BA\quad.
\label{(6)}
\ee
Since the $q$-factor $\chi\equiv q^{-g(A)g(B)}$ in \eq{(6)} is the same for any given $A$ and
$B$, irrespective of their order in the bracket, it follows that in contrast with \eq{(1)},
 $[A,B]_\chi$ and $[B,A]_\chi$, are not in general related to 
each other by a simple multiplicative factor. From \eq{(2)} and \eq{(3)} 
it can be 
seen that the graded bracket satisfies the following expansion identities
\be
[AB,C]_\chi=A[B,C]_\chi+q^{-g(B)g(C)}[A,C]_\chi B\quad,
\label{(7)}
\ee
\be
[A,BC]_\chi=[A,B]_\chi C+q^{-g(A)g(B)}B[A,C]_\chi\quad.
\label{(8)}
\ee
However it does not satisfy a generalized Jacobi identity of the type given 
in \eq{(4)}.

We also make use of the Gauss numbers $[r]_q$,
\be
[r]_q={{1-q^r}\over{1-q}}\quad,
\label{(9)}
\ee
\be
[r]_q!=[r]_q[r-1]_q[r-2]_q...[2]_q[1]_q\quad,\quad
{\hbox{supplemented \hskip5pt by}}
\quad 
[0]!=1\quad,
\label{(10)}
\ee
with definition \eq{(10)} holding only when $r$ is a non negative integer. 
Note that when $q^n=1$, 
our grading scheme becomes 
degenerate, so that in effect  the grading of an element is only defined 
modulo $n$. In this case we also have $[r]_q=0$ when $r$ modulo $n$ 
is zero $(r\neq0)$.

\section{Graded \Q-calculus}

To define the $q$-calculus algebra in the single $q$-variable case, we 
introduce a grade 1 $q$-variable $\theta$ and two
grade $-1$ $q$-derivatives ${\cal D}_L$ and ${\cal D}_R$. These are defined 
algebraically by the relations 
\be
[{\cal D}_L,\theta]_q:=1\quad,\quad
[\theta,{\cal D}_R]_q:=1\quad,
\label{(11)}
\ee
where until further notice we work with generic $q$, by which we mean 
$q$ not a root of unity. From a dimensional point of view, 
$[{\cal D}_L]=[{\cal D}_R]=[\theta]^{-1}$. By 
considering the relationship $[{\cal D}_L,[{\cal D}_R,\theta]_{q^{-1}}]=0$, or equally 
$[{\cal D}_R,[{\cal D}_L,\theta]_q]=0$, we establish using \eq{(4)} that 
\be
[[{\cal D}_L,{\cal D}_R]_{q^{-1}},\theta]=0\quad.
\label{(12)}
\ee
The simplest way to impose this condition is to
supplement definition \eq{(11)} by 
\be
[{\cal D}_L,{\cal D}_R]_{q^{-1}}=0\quad,
\label{(13)}
\ee
which ensures that our algebra closes under bilinear $q$-brackets.
Using \eq{(3)} it follows that
\be
[{\cal D}_L,\theta^m]_{q^{m}}=[m]_q\theta^{m-1}=[\theta^m,{\cal D}_R]_{q^{m}}\quad.
\label{(13a)}
\ee
Other differential operators
${\cal D}_s$ satisfying $[{\cal D}_s,\theta]_{q^s}=c_s$, where $c_s$ is 
a complex number, can be defined in a way compatible with ${\cal D}_L$ and 
${\cal D}_R$. 
Including them 
would mean that algebraic closure
could only be achieved by making use of trilinear $q$-brackets. For this 
reason we exclude them from our present treatment, but they perhaps form the 
basis of an interesting extension along the lines of the paraoscillator 
generalizations of the ordinary bosonic and fermionic algebras.

We define left and right differentiation in such a way that they are induced 
by the graded
bracket. Let $f(\theta)$ be a function of $\theta$ which can be expanded as a 
power series with commuting complex coefficients $C_m$,
\be 
f(\theta)={\sum_{m=0}^\infty} C_m\theta^m\quad.
\label{(15)}
\ee 
Then left differentiation of $f(\theta)$ is induced by the grading 
 of ${\cal D}_L\,[(-1)]\,,\,\theta\,[(1)]$ and \eq{(6)} as follows
\be
\eqalign{
	\left({{df(\theta)}\over {d\theta}}\right)_L &
	\equiv[{\cal D}_L,f(\theta)]_\chi
	={\sum_{m=0}^\infty} C_m[{\cal D}_L,\theta^m]_\chi
	\cr
	&:={\sum_{m=0}^\infty} C_m[{\cal D}_L,\theta^m]_{q^m}
	={\sum_{m=1}^\infty} C_m[m]_q\theta^{m-1}\quad.
	\cr}
\label{(16)}
\ee 
Similarly right differentiation is defined by
\be
\left({{df(\theta)}\over {d\theta}}\right)_R \equiv [f(\theta),{\cal D}_R]_\chi:=\sq
C_m[m]_q\theta^{m-1}\quad.
\label{(17)}
\ee
For the reason given after \eq{(6)}, the definitions \eq{(16)} [\eq{(17)}] 
{\it require} that ${\cal D}_L$ $[{\cal D}_R]$ is placed at the left [right] in the 
$\chi$-bracket.

Comparing this with \eq{(16)} we see that left and right differentiation have 
the same effect, although the
associated algebraic operators ${\cal D}_L$ and ${\cal D}_R$ are different. In the $q=1$ 
case, corresponding to undeformed calculus, we have ${\cal D}_L=-{\cal D}_R$. Later we 
will establish the analogue of this result for generic $q$. 
More generally, we consider functions $f(\theta)$, the series expansions of 
which have graded, noncommuting, but still constant coefficients $A_m$,
\be 
f(\theta)={\sum_{m=0}^\infty} \theta^m A_m\quad.
\label{(18)}
\ee
We require that the coefficients $A_m$ have
bilinear commutation relations with $\theta$, ${\cal D}_L$ and ${\cal D}_R$.
Using the graded bracket to induce left differentiation we find, 
\be
\eqalign{
	\left({{df(\theta)}\over {d\theta}}\right)_L 
	&=\ss [{\cal D}_L,\theta^mA_m]_\chi\cr
	&=\ss[{\cal D}_L,\theta^m]_\chi A_m+\ss q^m\theta^m[{\cal D}_L,A_m]_\chi\cr
	&=\sq[m]_q\theta^{m-1}A_m+\ss q^m\theta^m[{\cal D}_L,A_m]_{q^{g(A_m)}}\quad.
	\cr}
\label{(19)}
\ee
This takes on the expected form if the second term vanishes, so we impose 
the following condition on $A_m$
\be
[{\cal D}_L,A_m]_{q^{g(A_m)}}=0\quad,
\label{(20)}
\ee
for all $m$. 
A similar treatment of right derivatives leads to
\be
[A_m,{\cal D}_R]_{q^{g(A_m)}}=0\quad.
\label{(21)}
\ee
For $g(A_m)\neq 0$ conditions \eq{(20)} and \eq{(21)} are not compatible since,
using the Jacobi identity \eq{(4)} for 
$\theta\,,\,A_m$, ${\cal D}_L$ and for $\theta\,,\,A_m$, and ${\cal D}_R$, it 
follows that
\be
[{\cal D}_L,A_m]_\gg=0,\hskip5pt \Rightarrow\hskip5pt
[A_m,\theta]_\gg=0,\hskip5pt\Rightarrow\hskip5pt[{\cal D}_R,A_m]_\gg=0\quad,
\label{(22)}
\ee
and the last one is different from \eq{(21)}.
This means that in general when dealing with functions of $\theta$ we must 
single out either left or right differentiation. 

Suppose we choose left 
differentiation. Then condition \eq{(20)} ensures that
the coefficients $A_m$ have commutation relations compatible with their 
grading in such a way that the graded bracket induces differentiation. Note 
that if the coefficients of the power series expansion of $f(\theta)$ satisfy 
\eq{(20)} then so will the coefficients of the power series expansion
of its left derivative. As an example  we consider a graded version of the
$q$-exponential \r{CP,GRA,MajidI},
defined by
\be
\eqalign{
	\exp_q(\theta A)=\ss {(\theta A)^m\over [m]_q!}\quad.\cr}
\label{(23)}
\ee
Then imposing 
\be
[{\cal D}_L,A]_{q^{g(A)}}=0\quad,
\label{(24)}
\ee
which corresponds to \eq{(20)}, the bracket induced derivative has the 
expected form,
\be
\eqalign{ 
	\left({{d \exp_q(\theta A)} \over {d\theta}}\right)_L
	&=[{\cal D}_L,\ss {(\theta A)^m\over [m]_q!}]_\chi
	\cr
	&=\ss {1\over [m]_q!}  [{\cal D}_L,(\theta A)^m]_{q^{m(1+g(A))}}
	\cr
	&=A\exp_q(\theta A)\quad.}
\label{(25)}
\ee

We now introduce left and right $q$-integration\r{ChryZu,KM}, 
defining these 
in a natural way so that up to a (in general non central)
integration constant term (suppressed) they invert the effect of 
$q$-differentiation. 
We do this by imposing
\be
\eqalign{
	[{\cal D}_L,\int (d\theta)_L \hskip5pt\theta^m]_\chi&:=\theta^m\quad,
	\cr
	[\int\theta^m\hskip5pt (d\theta)_R,\hskip5pt {\cal D}_R]_\chi&:=
        \theta^m\quad,
	\cr}
\label{(26)}
\ee
where the expressions in brackets $(d\theta)_L$ and $(d\theta)_R$ denote 
left and right integral measures, with the same dimensions as $\theta$.
Using \eq{(13a)}, these give
\be\eqalign{ 
	\int (d\theta)_L \hskip5pt \theta^m&={\theta^{m+1}\over[m+1]_q}\quad,
	\cr
	\int\theta^m\hskip5pt (d\theta)_R&={\theta^{m+1}\over[m+1]_q}\quad,\cr}
\label{(27)}
\ee
which can be extended to arbitrary functions $f(\theta)$ by linearity.
Selecting left differentiation, and imposing \eq{(24)} we can again use 
the $q$-exponential to provide a simple example,
\be 
\int (d\theta)_L\hskip5pt A\exp_q(\theta A)=\exp_q(\theta A)\quad,
\label{(31)}
\ee
which by \eq{(24)} is equivalent to
\be 
\int (d\theta)_L\hskip5pt \exp_q(\theta A)=A^{-1}\exp_q(\theta A)\quad,
\label{(32)}
\ee
when $A$ is invertible.

\section{Shift operators for functions of $\theta$}

Let $f(\theta)$ be a function of the form \eq{(18)} with no conditions on the
commutation properties of the coefficients
$A_m$. 
Defining a left shift as 
$\theta\mapsto\epsilon+\theta$ where 
$[\theta,\epsilon]_q=0$
(a relation that will be justified later on), 
we now search for an invertible left shift operator $G_L$ such that
\be 
G_L f(\theta)G_L^{-1}=f(\epsilon+\theta)\quad.
\label{(34)}
\ee
By substituting into this the power series form of $\ff$ \eq{(18)} we find
\be
\eqalign{
	f(\epsilon+\theta)&
	=\ss G_L\theta^m A_mG_L^{-1}\cr
	&=\ss G_L\theta G_L^{-1}G_L\theta G_L^{-1}...G_L\theta G_L^{-1}
	G_LA_mG_L^{-1}
	\cr
	&=\ss (\epsilon+\theta)^m A_m\quad.\cr}
\label{(35)}
\ee
Thus any solution to $G_L\theta G_L^{-1}=\epsilon+\theta$, fulfilling the 
condition  $G_LA_mG_L^{-1}=A_m$, solves \eq{(34)} as well.
The solution in the undeformed case suggests that the operator we are 
looking for is some sort
of exponentiation of the degree zero operator $\ed$. 
For this reason we start with the power series
\be 
G_L=\ss C_m (\ed)^m\quad,
\label{(36)}
\ee
where the coefficients $C_m$ are complex numbers to be determined. 
The condition 
$G_LA_mG_L^{-1}=A_m$ now reduces to
\be
[\ed,A_m]=0\quad.
\label{(37)}
\ee
This should be interpreted as a condition on $\epsilon$ rather than on $A_m$, 
since it imposes upon
$\epsilon$ commutation properties identical to those of $\theta$ 
(and establishes that, as is the case for $\epsilon Q$ in supersymmetry, 
$\epsilon {\cal D}_L$ is commuting).
To find the coefficients $C_m$ we substitute \eq{(36)} 
into $(\epsilon+\theta)G_L=G_L\theta$ 
as follows.
\be
\eqalign{
	(\epsilon+\theta){\sum_{n=0}^\infty} C_n (\ed)^n&
	=\ss C_m (\ed)^m\theta
	\cr
	&=\ss (C_m\theta(\ed)^m+C_m[m]_{q^{-1}}\epsilon(\ed)^{m-1})\quad,
	\cr}
\label{(38)}
\ee
where $q^m\epsilon^m\theta=\theta\epsilon^m$ has been used. 
By equating coefficients of $\epsilon^m$ we find $C_m[m]_{q^{-1}}=C_{m-1}$. 
$G_L$ is only defined up to an overall multiplicative constant determined by 
$C_0$; as a group-like expression it is sensible to set $C_0=1$ in the 
expansion \eq{(46)}. 
This leads to
\be 
G_L=\ss{(\ed)^m\over{[m]_{q^{-1}}!}}
=\ss {{\epsilon^m {\cal D}_L^m}\over{[m]_q!}} =\exp_{q^{-1}}(\ed)\quad. 
\label{(39)}
\ee
It is well known\r{CP,GRA} 
that the inverse of the $q$-exponential exp$_qX$ is 
 exp$_{q^{-1}}(-X)$,
so the inverse of $G_L$ is
\be 
G_L^{-1}=\exp_q(-\ed)\quad.
\label{(40)}
\ee
Thus for any $f(\theta)$ we have the result 
\be 
f(\epsilon+\theta)=\exp_{q^{-1}}(\ed)f(\theta)\exp_q(-\ed)\quad.
\label{(41)}
\ee
Expression \eq{(39)} for $G_L$ agrees with the expression for the left 
shift $L_\epsilon=\exp_{q^{-1}}(\epsilon Q)$ in fractional supersymmetry 
(see sec. X) found in \r{AM} once ${\cal D}_L$ and $Q$ are identified.
In fact, results \r{AM,MajidI} equivalent to 
\be
\exp_{q^{-1}}(\ed)f(\theta)\zz=f(\epsilon+\theta)\zz\quad,
\label{(42)}
\ee
have been already established, 
but we believe that this is the first time an entirely 
algebraic form has been given.

For right shifts $\theta\rightarrow\theta +\eta$ where
$[\theta,\eta]_{q^{-1}}=0$ and hence
$\eta^m\theta=q^m\theta\eta^m$ there is an exactly analogous result. 
The right analogue of condition \eq{(37)} for the right derivation is
\be
[A_m, {\cal D}_R\eta]=0\quad.
\label{(43)}
\ee
Provided that this is satisfied we have $G_R=\exp_q(-{\cal D}_R\eta)$ and 
\be 
f(\theta+\eta)=\exp_q(-{\cal D}_R\eta)f(\theta)\exp_{q^{-1}}({\cal D}_R\eta)
\quad.
\label{(44)}
\ee
Correspondingly, $G_R$ may be seen equal to the right shift in 
fractional supersymmetry $R_\eta=\exp_q(\eta D)$ of \r{AM}, once ${\cal D}_R$ 
is identified with $-q^{-1}D$ (see \eq{(92)} below) and 
${\cal D}_R\eta=q\eta {\cal D}_R$
(which follows from $[\theta,\eta]_{q^{-1}}=0$) has been used.
Also $[\epsilon {\cal D}_L, {\cal D}_R\eta]=0$ so that, as expected, 
left and right shifts commute.

\section{Number operators in \Q-calculus}

We now seek a number operator $N$ satisfying
\be
[N,{\cal D}_L]=-{\cal D}_L\quad,\quad
[N,{\cal D}_R]=-{\cal D}_R\quad,\quad
[N,\theta]=\theta\quad.
\label{(45)}
\ee
To get the explicit form of $N$, we note that from \eq{(5)} and 
\eq{(45)} $g(N)=0$, 
which 
suggests that if $N$ exists, its power series expansion will have the form
\be
 N=\ss C_m\theta^m {\cal D}_L^m\quad,
 \label{(46)}
 \ee
where the $C_m$ are here complex numbers which we can find by solving 
\be
\eqalign{
	[N,\theta]&=\ss[C_m\theta^m {\cal D}_L^m,\theta]\cr
	&=\ss(C_m\theta^m[{\cal D}_L^m,\theta]_{q^m}+
	C_m q^m[\theta^m,\theta]_{q^{-m}}{\cal D}_L^m)
	\cr
	&=\ss([m]_qC_m\theta^m {\cal D}_L^{m-1}-C_m(1-q^m)\theta^{m+1}{\cal D}_L^m)\quad.\cr}
\label{(47)}
\ee
Equating coefficients gives $C_0$ undetermined, $C_1=1$, and for $m\geq2$,
\be
[m]_qC_m=(1-q)[m-1]_qC_{m-1}\quad.
\label{(48)}
\ee
So our result is 
\be 
N=C_0+\sq {(1-q)^{m-1}\over[m]_q} \theta^m {\cal D}_L^m\quad.
\label{(49)}
\ee

It is easy to verify that $[N,{\cal D}_L]=-{\cal D}_L$. $[N,{\cal D}_R]=-{\cal D}_R$ 
follows from the 
connection \eq{(60)} between ${\cal D}_L$ and ${\cal D}_R$, shortly to be derived. 
Note that we 
could just as well have chosen to expand $N$ using $\theta$ and ${\cal D}_R$. 
Power series expansions of $q^{rN}$ are also of interest. To construct
these start with the expansion
\be 
q^{rN}=\ss B_m\theta^m {\cal D}_L^m\quad,
\label{(50)}
\ee 
and determine the complex coefficients by solving
\be
\theta q^{r(N+1)}=q^{rN}\theta\quad.
\label{(51)}
\ee
Thus we have,
\be
\eqalign{
	\ssn q^rB_n\theta^{n+1} {\cal D}_L^n&=\ss B_m\theta^m {\cal D}_L^m\theta
	\cr
	&=\ss(q^m B_m\theta^{m+1} {\cal D}_L^m+B_m[m]_q\theta^m {\cal D}_L^{m-1})\quad.
	\cr}
\label{(52)}
\ee
Equating coefficients we find $B_0$ undetermined and
\be 
q^mB_m+B_{m+1}[m+1]_q=q^rB_m\quad,
\label{(53)}
\ee
which leads to
\be 
q^{rN}=B_0+B_0\sq{1\over[m]_q!}({\prod_{p=0}^{m-1}}(q^r-q^p))\theta^m {\cal D}_L^m
\quad.
\label{(54)}
\ee
Although $B_0$ is 
undetermined, it is related to our choice of $C_0$ in 
\eq{(49)}. 
The easiest way to find the correspondence between $B_0$ and $C_0$ is to make 
use once more of an arbitrary 
representation, in a basis with ${\cal D}_L\zz=0$. 
Then we have
\be 
N\zz=C_0\zz\quad,\quad 
q^{rN}\zz=B_0\zz=q^{rC_0}\zz\quad,
\label{(55)}
\ee
which gives
\be 
B_0=q^{rC_0}\quad.
\label{(56)}
\ee

Note that since 
we have only made use of representations to establish a connection 
between two normalizations the result \eq{(56)} is not representation 
dependent, 
{{\it i.e.}} it holds for the algebra itself. We can make immediate use of 
this result to relate ${\cal D}_L$ and ${\cal D}_R$. From \eq{(54)} and with 
$B_0=1$ $(C_0=0)$, 
\be 
q^N=1+(q-1)\theta {\cal D}_L=
[{\cal D}_L,\theta]
\quad,
\label{(57)}
\ee
and hence for nonnegative integer $r$ (for which \eq{(54)} terminates),
\be 
q^{rN}=({\cal D}_L\theta-\theta {\cal D}_L)^r=[{\cal D}_L,\theta]^r
\quad.\label{(58)}
\ee 
Using \eq{(57)} and \eq{(51)} we get
\be
[q^{-N}{\cal D}_L,\theta]_{q^{-1}}=q^{-N}[{\cal D}_L,\theta]=1\quad,\label{(59)}
\ee
and comparing this to \eq{(11)} we find,
\be 
{\cal D}_R=-q^{-(1+N)}{\cal D}_L\quad.
\label{(60)}
\ee
Thus $[N,{\cal D}_L]=-{\cal D}_L\Rightarrow[N,{\cal D}_R]=-{\cal D}_R$ 
as claimed in \eq{(45)}. A similar 
treatment in which $q^{rN}$ is expanded using ${\cal D}_R$ 
instead of ${\cal D}_L$ leads to 
a result corresponding to \eq{(58)}. For any nonnegative integer $r$, this is
\be 
q^{-rN}=q^r(\theta {\cal D}_R-{\cal D}_R\theta)^r=
q^r[\theta,{\cal D}_R]^r
\quad.
\label{(61)}
\ee
Note that when expressed in terms of ${\cal D}_L$, the series corresponding to these 
$q^{-rN}$ do not
terminate. Using \eq{(57)} and \eq{(61)} 
we find the following algebra identities,
\be
\eqalign{
	{\cal D}_L\theta=[N+1]_q\quad,&\quad \theta {\cal D}_L=[N]_q\quad,\cr
	{\cal D}_R\theta=-q^{-1}[N+1]_{q^{-1}}\quad,&\quad
	\theta {\cal D}_R=-q^{-1}[N]_{q^{-1}}\quad.\cr}
\label{(62)}
\ee

\section { The connection to \Q-deformed bosons.}

The results of section V readily indicate how to relate the $q$-calculus to 
the algebra of a single $q$-deformed bosonic oscillator \r{QQ}. 
We begin by writing
\be \theta=f_1(N) a_+ \quad; \quad {{\cal D}}_L=f_2(N) q^{N/2}a_- \quad .
\label{(6.1)} \ee 
\noindent It follows now from \eq{(11)} that
\be [a_- \, , \, a_+]_{q^{1/2}}=q^{-N/2} \label{(6.2)} \ee 
\noindent provided that
\be  f_2(N)f_1(N+1)=1. \label{(6.3)} \ee 
\noindent It follows similarly from \eq{(58)} and \eq{(6.3)} that
\be [a_- \, , \, a_+]_{q^{-1/2}}=q^{N/2}\quad. \label{(6.4)} \ee 
\noindent Eqs. \eq{(6.2)} and \eq{(6.3)} are familiar as the defining equations
of the $q$-deformed bosonic oscillator \r{AC,Macfarlane,Biedenharn}. However care must be taken in 
comparing $a_{\pm}$ with the corresponding creation and annihilation operators
$a$ and $a^{\dagger}$.

Consider first for real $q$ the Fock space spanned by the kets $|m\rangle \, ,
\, m=0,1.2, \dots $, where the identifications $a_-=a \, , \,a_+=a^{\dagger}$,
are straightforward with the dagger implying hermitian conjugation in 
the usual way. Here $a|0\rangle =0$,  
$N|r\rangle =r|r\rangle \, $, and
\be\langle m+1 |a^{\dagger}|m\rangle =[[m+1]]_q^{1/2} \quad ,
\quad \langle m-1 |a|m\rangle =[[m]]_q^{1/2} \quad , \label{(6.5)} \ee 
\noindent where the notation
\be 
[[x]]_q={{q^{x/2}-q^{-x/2}} \over {q^{1/2}-q^{-1/2}}} =q^{(1-x)/2} [x]_q
\quad,
\label{(6.6)} 
\ee 
\noindent has been used to display the results in their usual form. 
If we make the choice
\be f_1(N)=[N]_q^{1/2}q^{{(N-1)/4}} \quad ,  \label{(6.7)} \ee 
\noindent then \eq{(6.5)} and \eq{(6.1)} yield the following representations of 
$\theta$ and ${{\cal D}}_L$
\be\langle m+1 |\theta|m\rangle =[m+1]_q \quad , \quad 
\langle m-1 |{{\cal D}}_L|m\rangle =1 \quad , \label{(6.8)} \ee 
\noindent both defining real matrices for real $q$.
It can be checked that (the first entry of) \eq{(11)} and \eq{(58)} are 
satisfied. 

We turn later to the situation when $q$ is a 
primitive $n$-th root of unity, $q={{\exp}}{2\pi i\over n}$, but it is worth 
remarking already at this
point that \eq{(6.5)} describes a representation of \eq{(6.2)} and \eq{(6.4)} 
in a positive definite Hilbert space in which 
$a^{\dagger}$ is indeed the hermitian conjugate of $a$, this being true because
$q^{1/2}$ occurs in \eq{(6.2)} and \eq{(6.4)} as the deformation parameter, rather than
$q$ itself.

We may also use our work on the $q$-calculus to derive results that hold in the
deformed boson context. For example, in the case of \eq{(49)} with $C_0=0$
(which follows our previous choice of $B_0=1$), we may use \eq{(6.1)} and \eq{(6.3)} to
obtain 
\be N=\sum_{m=1}^\infty {{(1-q)^m} \over {1-q^m}} (a_+)^m (q^{N/2} a_-)^m 
\quad . \label{(6.9)} \ee 
As expected this implies $ [N \, , a_{\pm}] =\pm a_{\pm} \quad$.
\vskip 25pt

\section{\Q-calculus at \Q\ a root of unity}

In the previous sections we have been working with generic $q$, 
\ie\ with the restriction $q^n\ne 1$.
When $q^n=1$, our $q$ calculus takes on an specially interesting 
form, to which we will devote the rest of this paper. 
Our results will be valid for any primitive $n$-th root of unity, but we use 
$q=e^{{2\pi i\over n}}$ in all of our examples. From \eq{(11)} follows a 
result which holds for any $q$
\be
\left[{\cal D}_L,{\theta^m\over[m]_q!}\right]_{q^m}=
{\theta^{m-1}\over[m-1]_q!}=
\left[{\theta^m\over[m]_q!},{\cal D}_R\right]_{q^m}
\quad,\quad
\hbox{for\hskip5pt positive\hskip5pt}m\quad.
\label{(69)}
\ee
We want to know what happens to this in the $q\rightarrow\qq$ limit. We take 
this limit along the $\vert q\vert=1$ circle.
Taking the limit in a different direction would alter the reality properties of
$z$ as defined in \eq{(74)} below, but this could easily be corrected for by introducing a phase
factor into the definition.  
In taking this limit, difficulties first arise when
$m=n$, because then $[n]_q!=0$ $(n\ne 1)$. 
To retain \eq{(69)} in this case its LHS and RHS must remain 
finite and nonzero. For $n\ne 1$, this can be achieved by requiring that 
${\theta^n\over[n]_q!}$ remains finite and nonzero when $q\rightarrow\qq$.
This can only be true if $\theta^n$=0 when $q=\qq$.
This is an extra condition
on the q-calculus, and if we are to build a consistent framework, it must be 
preserved by left and right shifts. Thus for a left shift 
$\theta\mapsto\epsilon+\theta$, with $[\theta,\epsilon]_q=0$, we require
\be 
\ql(\epsilon+\theta)^n=\ql\snn\epsilon^m\theta^{n-m}{[n]_q!\over[n-m]_q!
[m]_q!}=0\quad.
\label{(70)}
\ee 
Since 
\be
{[n]_q!\over[n-m]_q![m]_q!}=0
\label{(71)}
\ee
for $q=$exp$({2\pi i\over n})$ and  $0<m<n$, this is 
equivalent to requiring $\epsilon^n=0$ when $q=\qq$.
Likewise, preservation of $\theta^n$=0 by right shifts imposes the condition
$\eta^n=0$.  Now, we note that for $\vert q\vert=1$, and $q$
not a root of unity, we have under complex conjugation,
\be
\overline{[n]_q!}=q^{-{1\over2}n(n-1)}[n]_q!\quad,
\label{(72)}
\ee
so that in the $q\rightarrow\qq$ limit
\be
\overline{[n]_q!}=-(-1)^n[n]_q!\quad.
\label{(73)}
\ee
Looking at \eq{(73)} it is clear that if we define
\be 
z=\ql  {\theta^n\over[n]_q!}\quad,
\label{(74)}
\ee
then for a realization in which $q^m\theta$ is real for some integer $m$
(the simplest hypothesis is to take $\theta$ itself real, see section XI and 
appendix A), 
$z$ is real when $n$ is odd, and imaginary when $n$ is even. 
By using the identities
\be
\eqalign{
        \ql\left({[mn]_q\over[n]_q}\right)&
        =\ql\left({1-q^{mn}\over1-q^n}\right)\cr
        &=\ql(1+q^n+q^{2n}+...+q^{(m-1)n})\cr 
        &=m\quad,\cr}
\label{(75)}
\ee
and for $0<m<n$,
\be
\eqalign{
        \ql\left({[rn+m]_q\over[m]_q}\right)&=\ql
        \left({1-q^{rn+m}\over1-q^m}\right)
        =1\quad,\cr}
\label{(76)}
\ee
we have, for $0\leq p<n$,
\be
\ql\left({\theta^{rn+p}\over[rn+p]_q!}\right)={\theta^p\over[p]_q!}
{z^r\over r!}\quad.
\label{(77)}
\ee
We introduced $z$ to deal with the difficulties encountered in the $q\to\qq$ 
limit of \eq{(69)} at $m=n$. We run into similar problems for all $m>n$, and the 
importance of \eq{(77)} is that it tells us that all of these can also be handled 
in terms of $z$. 

To investigate further the properties of
$z$, and to see how it fits into $q$-calculus, consider now 
the identity
\be 
[{\cal D}_L,...[{\cal D}_L,[{\cal D}_L,[{\cal D}_L,
{\theta^n\over[n]_q!}]_{q^n}]_{q^{n-1}}]_{q^{n-2}}]
...]_q=1\quad,
\label{(78)}
\ee
valid for generic $q$. 
In the limit as $q\to\qq$ this 
conveniently reduces to the commutator
\be
[{\cal D}_L^n,z]=1\quad([z,{\cal D}_R^n]=1)\quad.
\label{(79)}
\ee
So by defining 
\be
\dz:={\cal D}_L^n\quad  (=-{\cal D}_R^n)\quad,
\label{(80)}
\ee
we have
\be
[\dz,z]=1\quad.
\label{(81)}
\ee
Thus we see that the algebra associated with ordinary calculus emerges as a 
natural component of the $q$-calculus algebra in the $q\to\qq$ limit.
It is clear that there are alternative ways of keeping
$[{\cal D}_L^n,{\theta^n\over [n]_q!}]$ 
finite, corresponding to a different distribution of the $[n]_q!$ factor
between the definitions of $z$ and $\dz$
(for instance, $z={\theta^n\over [n]_q!^\alpha}\,,\,\partial_z=
{{\cal D}_L^n\over [n]_q!^{1-\alpha}}$). 
However, our choice $(\alpha=1)$ is the only one
which preserves the generic $q$ form of \eq{(69)}, and also the only one for which 
${\cal D}_L$ and ${\cal D}_R$ are not
nilpotent. The latter point is important because it means that we can define 
$q$-integration so that it inverts
the effect of $q$-differentiation (see section IX). 
We also  recall that when $q^n=1$ the definition of the grading of products 
as given by \eq{(5)} reduces to $g(XY)=(g(X)+g(Y))$mod $n$,
so that $g(z)=0$. Then from \eq{(81)}, $g(\dz)=0$ as well.
Let us now give a full set of relations for  $q$-calculus in the $q\to\qq$ 
limit: 
\be
\eqalign{
        \left({{d\theta} \over {d\theta}}\right)_L &:=
        [{\cal D}_L,\theta]_q=1\quad,
        \cr
        \left({dz\over {d\theta}}\right)_L &:=
        [{\cal D}_L,z]={\theta^{n-1}\over[n-1]_q!}
        \quad,\cr
        \left({{d\theta}\over {d\theta}}\right)_R &:=
        [\theta,{\cal D}_R]_q=1\quad,
        \cr
        \left({dz\over{d\theta}}\right)_R &:=
        [z,{\cal D}_R]={\theta^{n-1}\over[n-1]_q!}
        \quad,
        \cr
        \left({{\partial \theta} \over {\partial z}}\right) &:=
        [\dz,\theta]
        =[{\cal D}_L,...[{\cal D}_L,[{\cal D}_L,[{\cal D}_L,\theta]_q]_1
        ]_{q^{-1}}...]_{q^{2}}=0\quad,
        \cr
        \left({{\partial z} \over{\partial z}}\right) &:=[\dz,z]=1\quad.
        \cr}
\label{(82)}\ee
Since ${{\partial \theta}\over {\partial z}} =0$ and 
$\left({dz\over{d\theta}}\right)_L \neq0$, it is necessary to interpret 
$\left({\partial \over{\partial z}}\right)$ as a
partial derivative, and $\left( {d \over{d\theta}}\right)_L$
as a total derivative, a result which we took into account
when choosing our notation. 
We can also introduce  a left partial derivation with respect
to $\theta$, $\pt$, and a corresponding algebraic operator $\partial_\theta
\equiv \partial_L$. 
This partial derivative satisfies
\be
\eqalign{
        \pt\theta&:=[\partial_\theta,\theta]_q=1\quad,\cr
        \pt z&:=[\partial_\theta,z]=0\quad.\cr}
\label{(83)}
\ee
Using this we obtain a chain rule expansion of the total derivative,
\be
\eqalign{
        \left({d \over {d\theta}}\right)_L&=\left({{d\theta} \over 
        {d\theta}} \right)_L  \,
        \pt+\left({dz \over{d\theta}}\right)_L 
        {\partial \over {\partial z}}
        \cr
        &=\pt+\TT {\partial \over {\partial z}}
        \cr}
\label{(84)}\ee
or, in terms of the algebraic operators of fractional superspace
\be 
{\cal D}_L=\partial_\theta+\TT\dz\quad.
\label{(85)}
\ee

To go further we will need to make use of an identity \r{DurandII}, 
which we derive as follows,
\be
\eqalign{
        1=[{\cal D}_L^n,z]&={\cal D}_L^{n-1}[{\cal D}_L,z]+[{\cal D}_L^{n-1},z]{\cal D}_L
        \cr
        &=
        {\cal D}_L^{n-1}[{\cal D}_L,z]+{\cal D}_L^{n-2}[{\cal D}_L,z]{\cal D}_L+
        {\cal D}_L^{n-3}[{\cal D}_L,z]{\cal D}_L^2+...[{\cal D}_L,z]
        {\cal D}_L^{n-1}
        \cr
        &=
        {1\over[n-1]!}\sn {\cal D}_L^m\theta^{n-1}{\cal D}_L^{n-1-m}
        \cr
        &=
        {1\over[n-1]!}\sn \partial_\theta^m\theta^{n-1}\partial_\theta^{n-1-m}
        \quad.
        \cr}
\label{(86)}
\ee
Here we have used \eq{(82)} 
and $\theta^n=0$. By substituting our expansion \eq{(85)} 
into \eq{(80)}, and using \eq{(86)} as follows
\be
\eqalign{
        \dz&={\cal D}_L^n=\partial_\theta^n+{1\over[n-1]_q!}
        (\sn\partial_\theta^m\theta^{n-1}\partial_\theta^{n-1-m})\dz\cr
        &=\partial_\theta^n+\dz\quad,
        \cr}
\label{(87)}
\ee
we obtain the condition
\be
\partial_\theta^n=0\qquad
\hbox{or\hskip5pt equivalently}
\qquad
{\partial^n\over\partial^n\theta}=0\quad.
\label{(88)}
\ee
There are analogous results for right derivatives
$(\partial_{\theta R}\equiv\delta_\theta)$ and we summarize these below.
\be
\eqalign{
        \left({d \over {d\theta}}\right)_R&={\delta\over\delta\theta}-\TT
        {\partial \over {\partial z}}\quad,
        \cr
        {\cal D}_R&=\delta_\theta-\TT\dz\quad.
        \cr}
\label{(89)}
\ee
Here the sign change on the second term arises because the algebraic element 
associated with right
differentiation of $z$ is $-\dz$, \ie\ $[f(z),-\dz]= 
{\partial \over {\partial z}} f(z)$. 
Also 
\be
[\theta,\delta_\theta]_q=1\quad\hbox{and}\quad
\delta_\theta^n=0\quad,
\label{(90)}
\ee 
and from \eq{(13)} we find that
\be
[\partial_\theta,\delta_\theta]_{q^{-1}}=0\quad.
\label{(91)}
\ee

The $q\to1$ limit is special, because when $q=1$, 
$[1]_q=1$ remains nonzero,
so that in this case it is not necessary to set $\theta=0$. It then follows from the
definitions \eq{(74)} and \eq{(80)} that for $q=1$ we can identify $\theta=z$
and $\partial_\theta=\dz$, and thus the
algebra can be described solely in terms of $z$ and $\partial_z$, as expected 
for the undeformed case. 
This resolves a problem encountered in previous work 
on generalized Grassmann calculus \r{DurandI,DurandII}, 
in which the $n=1$ case appeared 
to contain only the trivial relation $\theta=0$, so that the undeformed case 
could only be recovered through a $n\to\infty$ limit.

The above results are of course directly related to ordinary/fractional 
supersymmetry. 
To see this define
\be 
Q={\cal D}_L\quad,\quad
D=-q{\cal D}_R\quad,
\label{(92)}
\ee
and
\be 
t=\left\{\eqalign {& z\hskip5pt\hbox{for\hskip5pt} n
\hbox{\hskip5pt odd,}
\cr
 & iz\hskip5pt\hbox{for\hskip5pt} n
\hbox{\hskip5pt even.}\cr}\right.
\label{(93)}
\ee 
Then from \eq{(13)},
\be
[Q,D]_{q^{-1}}=0\quad,
\label{(94)}
\ee
and from \eq{(80)},
\be 
Q^n=\dz=
\left\{\eqalign {& \partial_t\hskip5pt\hbox{for\hskip5pt} n
\hbox{\hskip5pt odd,}
\cr
& i\partial_t\hskip5pt\hbox{for\hskip5pt} n
\hbox{\hskip5pt even.}\cr}\right.
\label{(95)}
\ee
and 
\be 
D^n=-(-)^n\dz=\left\{\eqalign {& \partial_t\hskip20pt\hbox{for\hskip5pt} n
\hbox{\hskip5pt odd,}\cr
& -i\partial_t\hskip5pt\hbox{for\hskip5pt} n
\hbox{\hskip5pt even.}\cr}\right.\label{(96)}\ee
When $n=2$, these relationships identify $Q$ and $D$, respectively, 
as the supercharge and 
covariant derivative from one-dimensional supersymmetry, 
and eqs. \eq{(85)}, \eq{(89)} give their explicit 
form in terms of the superspace algebra. 
For higher $n$ they provide the corresponding operators in fractional
superspace. Thus the form of these operators, and the nice properties with 
which they are associated, all follow from the underlying total derivatives of 
$q$-calculus at roots of unity.

Our previous results can all be expressed in terms of $Q$ and $D$, and so 
related to ordinary/fractional supersymmetry. For example, from \eq{(60)} 
we have
\be 
D=q^{-N}Q\quad,\label{(97)}
\ee
where
\be  
[N,\theta]=\theta\quad,\quad
[N,D]=-D\quad,\quad
[N,Q]=-Q\quad,
\label{(98)}
\ee
which we believe to be a new result. We now look at what happens to the 
structures which we derived earlier when $q\to\qq$ .

\section{Functions of $\theta$ when \Q\ is a root of unity}

Eq. \eq{(77)} shows that, when $q\to\qq$, 
any function $f(\theta)$ defined by a positive power series expansion
takes in general on the `fractional superfield' form $f(z,\theta)$. 
As a specific example we look again at the graded 
$q$-exponential, with $[{\cal D}_L,A]_{q^{g(A)}}=0$, 
so that left differentiation is induced by the grading. 
It is convenient to define the bosonic element
\be
A_z=(-1)^{(n-1)g(A)}A^n,
\label{(116.5)}
\ee 
and we also define the cut off $q$-exponential to be
\be 
\exp_{q,c}(\theta A)=\sn{(\theta A)^m\over[m]_q!}\quad,
\label{(117)}
\ee
where the subindex $c$ indicates that the sum is cut off after $n$ terms.
Using this we find that in the $q\to\qq$ limit, the $q$-exponential
decomposes as follows.
\be 
\eqalign{
        \ql\exp_q(\theta A)&=\ql\ss{(\theta A)^m\over[m]_q!}
        \cr
        &=\ql\ss{(\theta^m A^m)\over[m]_q!}q^{{1\over2}m(m-1)g(A)}
        \cr
        &=\exp(zA_z)\sn{(\theta A)^m\over[m]_q!}
        \cr
        &=\exp(zA_z)\exp_{q,c}(\theta A)\quad.
        \cr}
\label{(118)}
\ee
If we define, 
\be 
\eqalign{
        {\sinh}_q(\theta A)=\ss {(\theta A)^{2m+1}\over [2m+1]_q!}\quad,\quad
        {\cosh}_q(\theta A)=\ss {(\theta A)^{2m}\over [2m]_q!}\quad,}
\label{(119)}
\ee
so that
\be  
{\cosh}_q(\theta A)+{\sinh}_q(\theta A)=\exp_q(\theta A)\quad,
\label{(120)}
\ee
then in the $q\to\qq$ limit ${\sinh}_q(\theta A)$ and 
${\cosh}_q(\theta A)$
decompose as follows,
\be 
\eqalign{
        \ql{\sinh}_q(\theta A)&=\exp(zA_z){\sinh}_{q,c}(\theta A)\quad,\cr
        \ql{\cosh}_q(\theta A)&=\exp(zA_z){\cosh}_{q,c}(\theta A)\quad.\cr
\label{(121)}
}\ee
Here ${\sinh}_{q,c}(\theta A)$ and ${\cosh}_{q,c}(\theta A)$ are,
as in \eq{(119)}, cut off after a finite number of terms, to leave only those 
terms in which $\theta$ is raised to a power of $n-1$ of less. Similar 
decompositions are seen in the $q\to\qq$ limit of other $q$-functions, 
and in particular for many of the hypergeometric functions discussed 
in \r{GRA}.

\section{Integration when \Q\ is a root of unity}

Following the approach of section III, we define $q$-integration
so that it inverts the effect of $q$-differentiation. 
To get its explicit form, we note that 
the total derivative of a general term is
\be 
\left({d\over{d\theta}}\right)_L
\left(z^r\theta^s\right)=\delta_{s,0}\hskip1pt r z^{r-1}
{\theta^{n-1}\over[n-1]_q!}+(1-\delta_{s,0})[s]_q z^r\theta^{s-1}\quad.
\label{(99)}\ee
This leads directly to the following definition for integration at $q$ a root 
of unity (integration constant suppressed), 
\be 
\int (d\theta)_L\hskip5pt \theta^s z^r=(1-\delta_{s,n-1})
{z^r\theta^{s+1}\over [s+1]_q}
+\delta_{s,n-1}[n-1]_q!{z^{r+1}\over(r+1)}\quad.
\label{(100)}
\ee
We can expand this integral in a way analogous to the chain rule expansion of 
the total derivative. This leads to 
\be 
\eqalign{
        \int (d\theta)_L
        &=\int\left(d\theta\hskip5pt+
        {\partial^{n-1} \over\partial^{n-1}\theta}dz \right)
        \cr
        &=\int d\theta\hskip5pt+
        {\partial^{n-1}\over\partial^{n-1}\theta}\int dz\quad,
        \cr}
\label{(101)}
\ee 
where $d\theta$ and $dz$ denote 
`partial' integration measures, \ie\
$\int d\theta$ and  $\int dz$ treat
$z$ and $\theta$ respectively as constants
(notice that both terms are dimensionally homogeneous since from 
\eq{(74)}
$[z]=[\theta]^n$).
These component integrals are defined by
\be 
\int dz\hskip5pt z^m={z^{m+1}\over m+1}\quad,
\label{(102)}
\ee
\be 
\int d\theta\hskip 5pt\theta^m=(1-\delta_{m,{n-1}})
{\theta^{m+1}\over[m+1]_q}\quad.
\label{(103)}
\ee
The second of these is just the integration rule for generalized Grassmann 
variables suggested in \r{FIK}.
This clearly differs from 
the Berezin integral used in ordinary supersymmetry, as well as from its fractional
analogue, which is derived below; in particular, the integration 
measure $d\theta$ in \eq{(103)} has the same dimensions as $\theta$.
We can obtain the integral analogue of \eq{(80)} 
by performing a succession of $n$ 
integrations as follows, 
\be 
\eqalign{
        \int (d^n\theta)_L\hskip5pt\theta^s z^r
        &=
        \int (d^{s+1}\theta)_L\hskip5pt\theta^{n-1}{[s]_q!\over[n-1]_q!} z^r
        \cr
        &=
        \int (d^s\theta)_L\hskip5pt [s]_q!{z^{r+1}\over r+1}
        \cr
        &=
        \theta^s{z^{r+1}\over r+1}\quad,}
\label{(104)}
\ee
which implies that
\be 
\int (d^n\theta)_L=\int dz.
\label{(105)}
\ee
Similarly, from \eq{(103)} we obtain after a succession of $n$ integrations 
the integral analogue of \eq{(88)},
\be 
\int d^n \theta=0\quad.
\label{(106)}
\ee

There are similar results for right integrals.
In summary, for the $L,R$ integrals and derivatives
we have the following complementary results,
\be 
\int(d^n\theta)_L=\int  dz\quad,\quad
\int(d^n\theta)_R=-\int dz\quad,\quad
{\cal D}_L^n=\dz\quad,\quad
{\cal D}_R^n=-\dz\quad.
\label{(110)}
\ee
These have the following partial analogues:
\be 
\int d^n\theta=
\int\delta^n\theta=0=
\partial_\theta^n=
\delta^n_\theta\quad.
\label{(111)}
\ee

Let us now connect our results to previous work on Berezin 
integration. 
We begin by noting that in the undeformed case, the algebraic integral
\be
\int dt\hskip2pt f(t)\quad,
\label{(112)}
\ee
is related to an ordinary number,
\be
I(f,t_1,t_2)=\int_{t_1}^{t_2} dt\hskip2pt f(t)\quad,
\label{(113)}
\ee
by means of a sum over the (infinitesimal) contributions from the (continuous) 
eigenvalues of $t$ between $t_1$ and $t_2$. In field theories based
on actions we take the integral over all space, \ie\ over all real eigenvalues.
Usually this involves $t_1\to-\infty$, $t_2\to\infty$
and the numerical value of the integral is denoted by $I(f)$.
Similar methods have been used in the construction of numerical integrals based
on $q$-deformed algebraic integrals, see for example\r{KM,ChryZu}. 
An essential 
feature of such methods is that they involve a sum over the eigenvalues 
of $\theta$, or more strictly, of 
${\theta^m\over[m]_q!}$. 
To see how such a technique might be applied to our to our integral \eq{(101)},
we first write
\be
I(f)=\left(\int_S d\theta\hskip5pt+
        {\partial^{n-1}\over\partial^{n-1}\theta}\int_S dz
        \right)f(z,\theta)\quad.
\label{(114)}
\ee
Here the subscript $S$ 
indicates that the integral is to be taken over all 
space, \ie\ over all eigenvalues. The numerical value of this integral
is in general representation dependent, and may also depend on which particular integration
technique we use. 
However $\theta$ is nilpotent, so in any representation all of its eigenvalues
are zero. Consequently the first term makes no contribution to the numerical value of
the integral, and can be dropped.
If we now define a generalized Berezin integral by
\be
\int d\theta_{Ber}:={\partial^{n-1}\over\partial^{n-1}\theta}\quad,
\label{(115)}
\ee
we find that the (fractional) Berezin integral of a power series in
$\theta$ is given by 
\be
\int d\theta_{Ber}(c_0+c_1\theta+\ldots+c_{n-1}\theta^{n-1})=c_{n-1}[n-1]_q!
\quad \hbox{or} \quad
\int d\theta_{Ber}\theta^m=\delta_{n-1,m}[n-1]_q!\quad.
\label{bint}
\ee

Similar Berezin-like integrals were also considered in \r{BF,DurandII}.
In our framework we also 
obtain from \eq{(114)} the following integral measure on fractional
superspace,
\be
I(f)=\int_S dz d\theta_{Ber}f(z,\theta)\quad.
\label{(116)}
\ee
Using eq. \eq{(93)} to substitute $t$ for $z$ it is clear that up 
to an overall factor of $i$
this is equivalent, when $n=2$, to the familiar Berezin 
\r{Berezin} integral measure on one-dimensional superspace,
\be
I(f)=\int_S dt d\theta_{Ber}f(t,\theta)\quad,
\label{(116.1)}
\ee
in which case $\int d\theta_{Ber}\sim{\partial\over\partial\theta}$.
Thus the usual integral measure on superspace arises naturally out of our 
deformed
geometry based integral $\int(d\theta)_L$. This should be seen as an underlying
structure which gives the measure its specific and useful properties.
In particular, and in contrast to the standard Cartan differential measure, 
the Berezin integral element has dimensions
$[d\theta_{Ber}]=[\theta^{-1}]$, so that in general $ d\theta_{Ber}$ transforms 
with the inverse Jacobian under a coordinate change.
In the present $Z_n$-fractional case 
$[d\theta_{Ber}]=[\theta^{1-n}]$ and 
$d\theta_{Ber}=k^{1-n}d\theta'_{Ber}$ under the change $\theta=k\theta'$. 
Thus, our analysis above has enabled us to construct a fractional 
generalization of the usual Berezin integral and superspace measure
(the word measure being here understood in a formal sense).\par

\section{Shift and number operators when \Q\ is a root of unity}

We now examine the form taken by the operators derived in sections IV and V
when $q\to\qq$. Employing results from section VII, we find
\be 
\eqalign{
        G_L &
        = \ss {{\epsilon^m {\cal D}_L^m} \over {[m]_q !}}=
        \sum_{r=0}^{\infty} \sum_{p=0}^{n-1} {{\epsilon^p {\cal D}_L^p\epsilon^{rn}
        {\cal D}_L^{rn}} \over {[rn+p]_q !}} 
        =\sum_{r=0}^{\infty}\sum_{p=0}^{n-1}{{\epsilon^p {\cal D}_L^p}\over{[p]_q !}}
        {{z_\epsilon^r \partial_z^r}\over {r!}} 
        \cr
        &={\exp}(z_\epsilon \partial_z)\hskip2pt \sum_{p=0}^{n-1} 
        {{\epsilon^p {\cal D}_L^p}\over{[p]_q !}} 
        ={\exp}(z_\epsilon \partial_z) {\exp}_{q^{-1},c}(\epsilon {\cal D}_L)\quad,
        \cr} 
\label{(122)}\ee
\noindent with the truncated exponential as defined in \eq{(118)}. 
Also,
in \eq{(122)} the definition \eq{(80)} of $\partial_z$ has been employed and 
we  have been lead to make the definition
\be 
{z_\epsilon}=\ql  {\epsilon^n\over[n]_q!}\quad,
\label{(123)}
\ee
of the undeformed variable associated with $\epsilon$. 
It is 
independent of $z$ and has simply emerged naturally from our analysis.
We are thus able to conclude easily that the effect of a left shift on $z$ is
\be 
\eqalign{
        z\to G_L z G_L^{-1}&=
        \ql {(\epsilon+\theta)^n\over{[n]_q!}}
        =z+z_{\epsilon}+\sum_{m=1}^{n-1}
        {\epsilon^m\theta^{n-m}\over[m]_q![n-m]_q!}\quad.
        \cr}
\label{(124)}\ee

There is similar result for right shifts. On the other hand the fact that
$\theta^n=0$ causes the series expansion of the number operator 
to terminates. From \eq{(49)} this becomes
\be 
N=C_0+\snw {(1-q)^{m-1}\over[m]_q} \theta^m {\cal D}_L^m+(1-q)^{n-1}[n-1]_q!z\dz
\quad,
\label{(125)}
\ee
which by use of \eq{(84)} and the identity $(1-q)^{n-1}[n-1]_q!=n$, 
 which follows from \eq{(164)}, reduces to
\be 
N=C_0+\snw {(1-q)^{m-1}\over[m]_q} \theta^m \partial^m_\theta+nz\dz\quad.
\label{(126)}
\ee
Thus we can decompose the number operator into
\be 
N=N_c+nN_z\quad,
\label{(127)}
\ee
where $c$ indicates the cutting off of the sum after $n$ terms 
as in \eq{(117)} or \eq{(126)},
and $N_z=z\dz$ is just an ordinary number operator. The
complete set of commutation relations 
\be 
\eqalign{
        [N,z]&=nz\quad,\quad
        [N,\dz]=-n\dz\quad,
        \cr
        [N,\theta]&=\theta\quad,\quad
        [N,{\cal D}_L]=-{\cal D}_L\quad,
        \cr}
\label{(128)}\ee
follows from
\be 
\eqalign{
        [N_c,\theta]&=\theta\quad,\quad
        [N_c,\partial_\theta]=-\partial_\theta\quad,\quad
        [N_c,z]=0\quad,\quad
        [N_c,\dz]=0\quad,
        \cr
        [N_z,z]&=z\quad,\quad
        [N_z,\dz]=-\dz\quad,\quad
        [N_z,\theta]=0\quad,\quad
        [N_z,\partial_\theta]=0\quad.
        \cr}
\label{(129)}\ee
The easiest way to obtain the form of $q^{rN}$ when $q\to\qq$ is to
use \eq{(127)} the decomposition of $N$, which gives
\be 
q^{rN}=q^{{rN_c}+rnN_z}=q^{rN_c}q^{rnN_z}\quad,
\label{(130)}
\ee 
since from \eq{(129)} it follows that $[N_c,N_z]=0$.
Setting $C_0$=0 so that $B_0$=1, we have 
\be 
q^{rN}=q^{rnN_z}\sn{1\over[m]_q!}({\prod_{p=0}^{m-1}}(q^r-q^p))\theta^m 
{\cal D}_L^m
\quad,
\label{(131)}
\ee
which from \eq{(85)} reduces to
\be 
q^{rN}=q^{rnN_z}\sn{1\over[m]_q!}({\prod_{p=0}^{m-1}}(q^r-q^p))\theta^m
\partial^m_\theta\quad.
\label{(132)}
\ee
Note also that when $r$ is an integer, we have from \eq{(58)}
and \eq{(85)}
\be
q^{rN}=[{\cal D}_L,\theta]^r
=[\partial_\theta,\theta]^r\quad.            \label{132.1}
\ee
Now \eq{(132)} can also be written as
\be
q^{rN}=q^{rnN_z}[\partial_\theta,\theta]^r\quad,  \label{132.2}
\ee
and by comparing the last two equations with 
\eq{(130)},
we prove that for integer $r$
\be
q^{rnN_z}=1\quad,\quad
q^{rN_c}=[\partial_\theta,\theta]^r \quad,
\label{132.3}
\ee
so that $N_z$ has integer eigenvalues.

\section{Relation to \Q-deformed bosons when \Q\ is a root of unity}

We show here how, when $q={{\exp}}({2\pi i\over n})$, $q$-calculus
can be realized in terms of one ordinary boson and one $q$-deformed boson with
deformation parameter $q^{1/2}$. When $q={{\exp}}({2\pi i\over n})$ we can consistently
impose the condition $a^n=a^{\dagger n}=0$ on the  $q$-deformed bosonic algebra. 
Having done this it is justified to write $a_-=a $ and $a_+=
a^{\dagger}$ within the formalism of section VI, since in this case there are (finite dimensional)
matrix representations with the implied Hermiticity properties (see appendix C).
 Also the natural occurrence
of creation and annihilation operators suggests that the formalism is 
well-suited to applications in quantum mechanics.

Definitions analogous to \eq{(6.1)} will of course here be employed when 
$q={{\exp}}({2\pi i\over n})$. In fact we have
\be
\theta=f_1(N)a^{\dagger} \quad ,
 \quad \partial_\theta =f_2(N) \, q^{N/2} \,a \quad , \label{(11.1)} 
\ee 
\noindent 
where $f_1\,,\, f_2$ are given by \eq{(6.7)} and \eq{(6.3)}.
Note that the nilpotency of $a$ means that it should be related to
$\partial_\theta$ rather than ${\cal D}_L$.
The relation \eq{(6.3)} is still needed to ensure that the relations
\be
 q^N=\partial_\theta \theta -\theta \partial_\theta \quad , \quad
[\partial_\theta \, , \, \theta ]_q =1 \quad , \label{(11.2)} 
\ee 
\noindent are satisfied. If the choice \eq{(6.7)} for $f_1(N)$ is retained, then 
\eq{(6.8)} for $\theta$ does not yield a real matrix. The matrix is however 
related by  unitary equivalence to one that is real. 
One reaches this explicitly by
replacing the choice \eq{(6.7)} by the alternative one
\be 
f_1(N)=[[N]]_q^{1/2}=q^{(1-N)/4} \, [N]_q^{1/2} \quad , \label{(11.3)} 
\ee 
\noindent which, since $[[N]]_q^{1\over2}a^\dagger|j\rangle 
=[[j+1]]_q |j+1\rangle$,
gives rise to the representation
\be
 \langle j+1|\theta |j\rangle =[[j+1]]_q =q^{-j/2} [j+1]_q \quad , \label{(11.4)} 
\ee 
which is indeed real. Using \eq{(6.3)} for the new choices of 
$f_1,f_2$, we find
\be
 \langle j-1|\partial_\theta |j\rangle =q^{(j-1)/2}  \quad .\label{(11.5)} 
\ee 
\noindent It can be checked that \eq{(11.4)}-\eq{(11.5)} do indeed also satisfy \eq{(11.2)}. If
we temporarily denote the quantity that corresponds to the choice \eq{(11.3)} by
$\tilde \theta$ then using $f(N)\theta=\theta f(N+1)$, it is easy to find the relationship 
\be
 \tilde \theta=q^{(1-N)/2} \theta = \theta q^{-N/2}
=q^{N(1-N)/4} \, \theta q^{-N(1-N)/4} \quad \label{(11.6)} 
\ee 
\noindent
which exhibits the unitary equivalence. For many purposes it is simpler to
employ the original form of the representation that stems from \eq{(6.7)}. 
Further discussion of representations of the theory for $q$ a root of unity is
presented in appendix A. Note that if we represent $z$ and $\partial_z$ using 
an ordinary bosonic variable $b$ and its adjoint 
$b^{\dagger}$, $[b \, , \, b^{\dagger}]=1$, then we may write
\eq{(84)} and \eq{(85)} entirely in terms of the creation and annihilation 
operators of bosonic and $q$-deformed bosonic oscillators.
As one would expect 
the $q$-deformed oscillators at roots of unity have properties analogous
to those of the $q$-calculus operators. For a discussion see \r{QQ,RD1}.

\section{The connection between fractional supersymmetry 
and the braided line}

In developing this paper we approached $q$-calculus from a point of view in 
which it is seen primarily as a mathematical tool for use in fractional 
supersymmetry. An important observation, however, is that in the generic $q$ 
$(q^n\ne 1)$
case covered in sections III-IV our calculus corresponds to the simplest 
example of a braided calculus \r{MAJBOOK,MajidI}, 
that associated with the braided (or 
quantum) line\r{MAJBOOK}, though many of our results are new even in this
context, as for example the introduction of the algebraic shift operator.
Furthermore, to our knowledge, the calculus obtained
by taking 
the $q\to\qq$ limit, rather than by simply setting $q=\qq$, has 
not been considered  previously, and it is interesting to make use of our 
results to examine the extra structure which this approach uncovers. From the 
braided line perspective, the left shift
$\theta\to\epsilon+\theta$ 
is generated by the braided coproduct,
\be 
\theta\to\Delta\theta=\theta\otimes1+1\otimes\theta\quad,\quad
(=\epsilon+\theta)\quad,\quad
\label{(138)}
\ee
where
\be 
(A\otimes B)(C\otimes D)=q^{g(B)g(C)}AC\otimes BD\quad,
\label{(139)}
\ee 
so that
\be
(1\otimes\theta)(\theta\otimes1)=q\theta\otimes\theta\quad,\quad
(\theta\otimes1)(1\otimes\theta)=\theta\otimes\theta\quad.
\label{(140)}
\ee
Using \eq{(138)} we find
\be 
\Delta\theta^r=\sr{[r]_q!\over[m]_q![r-m]_q!}\theta^m\otimes\theta^{r-m}\quad.
\label{(141)}
\ee
There exist also a counit and antipode,
\be 
\eqalign{
        \varepsilon(\theta)=0\quad,\quad
        S(\theta^r)=q^{r(r-1)\over2}(-\theta)^r\quad,}
\label{(142)}
\ee
which satisfy the usual Hopf algebraic relations so long as the braiding is 
remembered. Now, when $q\to\qq$, eqs. \eq{(140)} and \eq{(141)}
ensure that if $\theta^n=0$ 
then $\Delta\theta^n=0$. From \eq{(74)} and \eq{(77)} 
we know that in this limit a full 
description of the braided line requires the introduction of an extra 
algebraic element $z$. 
This provides us with a natural extension of 
the braided Hopf algebra associated with the 
anyonic line \r{MajidII,MAJBOOK}.
Using \eq{(74)} and \eq{(141)} we obtain the  coproduct of $z$, as 
well as its counit and antipode,
\be 
\eqalign{
        \Delta z&=z\otimes 1+1\otimes z+
        \snw{1\over[m]_q![n-m]_q!}\theta^m\otimes\theta^{n-m}\quad,
        \cr
        \varepsilon(z)&=0\quad,\quad
        S(z)=-z\quad.\cr}
\label{(143)}
\ee

This result is interesting, because it means that while $z$ and $\dz$ satisfy 
the algebra associated with ordinary calculus, $z$ does not have 
a primitive braided Hopf structure. 
In our treatment of the $q$-calculus algebra we were able to use 
partial derivatives to perform a separation of this into independent $z$ and 
$\theta$ algebras. 
When we look at the larger Hopf algebraic structure, we see
that since $\theta$ appears in the coproduct of $z$, we can perform no such 
separation, and hence this cannot be regarded as a composite entity. 
This non-primitive braided Hopf structure is fundamental to our view 
of supersymmetry and fractional supersymmetry. 
Setting $n=2$ and using \eq{(93)} to relate $z$
and $t$ we see that even in ordinary one-dimensional 
supersymmetry the coalgebra structure for 
the time variable is non-primitive, since it has coproduct
\be 
\Delta t=t\otimes1+1\otimes t+i\theta\otimes\theta\quad.
\label{(144)}
\ee
It is usual to regard supersymmetry as a symmetry between the odd and even 
sectors of a superspace.
More precisely \r{AA}, 
rigid supersymmetry may be regarded as the result of centrally 
extending the initially Abelian odd translation group by the ordinary (even) 
spacetime translations (here, just time).
Fractional supersymmetry can be similarly described \r{AM}.
The present work provides us with
a new geometric interpretation of supersymmetry
\r{US}, which applies equally in the fractional case \r{Z3}. 
We have seen that at $q$ a root of unity we can conveniently describe the 
braided line in terms a grade 1 variable $\theta$ and a grade 0 variable
$z$. 
Under a translation $\theta\mapsto\epsilon+\theta$ along this line, $z$ 
transforms as in \eq{(124)}, which in terms of $t$ is just the 
ordinary/fractional supersymmetry transformation. 
We may now identify the one-dimensional fractional 
superspace of order $n$ with the braided line when $q$ is
an $n$-th root of unity. We can then further identify a fractional 
supersymmetry transformation as a shift along this braided line. Thus 
fractional supersymmetry is no more than translational 
invariance along the braided line at $q$ an $n$-th root of unity, the $n=2$ 
case corresponding to ordinary supersymmetry \r{US}.

\section{Concluding remarks}

Interpreting our work in the light of the last section we can say that when 
$q$ is a root of unity the braided Hopf algebra associated with the braided 
line is most conveniently described by two variables, one of which satisfies 
the algebra associated with ordinary calculus but has 
non-primitive Hopf structure. 
The analysis remains substantially unchanged when we move from one-dimensional 
to $D$-dimensional FSUSY, although there are some extra subtleties.
Moreover, the above is only the simplest example of a much more 
general result. For example it is possible to perform a similar decomposition 
of all $sl_q(n)$ quantum hyperplanes when $q\to\qq$. In this limit the quantum
hyperplane is most conveniently described using twice as many variables as 
usual, the extra variables being those associated with an undeformed plane.
We would expect to see similar phenomena in a wide range of quantum and braided
groups, the general point being that to take the limit of the generic $q$ 
case when $q\to\qq$ we need to introduce new variables, and that these provide
the Hopf algebra with extra structure.
We shall come back to these and other points in the future.

\vskip20pt\noindent
\acknowledgements
\vskip10pt
This paper describes research supported in part by E.P.S.R.C and P.P.A.R.C 
(UK) and CICYT (Spain).
J.C.P.B. wishes to acknowledge an FPI grant
from the Spanish Ministry of Education and Science and the CSIC.

\appendix
\section{}

Here we study the passage to the limit in which $q={{\exp}}{2\pi i/n}$
of the matrix representations of the $q$-calculus. Since ${{\cal D}}_R$ is
related directly to ${{\cal D}}_L$, we omit details in relation to it, for they
are easily deduced from the results given below. We represent $\theta$ and 
${{\cal D}}_L$ in a vector space ${{\cal V}}$ spanned by the kets $|m\rangle ,
m=0,1,2, \dots$, where $|0\rangle $ is such that ${{\cal D}}_L|0\rangle =0$ and
$N|m\rangle =m|m\rangle $. It is simplest to pass to the limiting case of 
interest, using the representation \eq{(6.1)} corresponding to the choice \eq{(6.7)}
of $f_1(N)$ that yields the matrices \eq{(6.8)} for $\theta$ and ${{\cal D}}_L$.
When studying the situation that obtains when $q={{\exp}}{2\pi i/n}$ it is
usual to employ only finite dimensional representations, since the content of
matrices like those of \eq{(6.8)} somehow `repeats' after $n$-terms. However,
the extra structure discussed in this paper emerges when no such 
`simplification' is made.  Using 
$\theta|m\rangle=[m+1]_q|m+1\rangle$ (eq. \eq{(6.8)}) for generic $q$ gives 
\be
 {{\theta^n} \over {[n]_q!}} |m\rangle =
{{[m+n]_q!} \over {[m]_q! \, [n]_q!}} |m+n\rangle \quad . \label{(A.1)} 
\ee 
\noindent
Setting $m=rn+p$, and making careful use of \eq{(74)}-\eq{(76)} to pass to our 
limiting situation, converts \eq{(A.1)} into the form
\be
z|rn+p\rangle =(r+1) |(r+1)n+p\rangle \quad , \label{(A.2)} 
\ee 
\noindent for all integers $r$ and $p \in \{ 0,1,2, \dots ,(n-1) \}$. 
Using ${\cal D}_L|m\rangle=|m-1\rangle$ (eq. \eq{(6.8)}) the 
corresponding result for $\partial_z =({{\cal D}}_L)^n$ is
\be 
\partial_z|rn+p\rangle = |(r-1)n+p\rangle  \quad . \label{(A.3)} 
\ee 
\noindent These results are in agreement with \eq{(81)}. 
If in an analogous fashion we now use $\tilde\theta$, the real
representation of $\theta$ from section XI, we find from \eq{(11.3)} and \eq{(A.1)}
that
\be
{{{\tilde\theta}^n} \over {[n]_q!}} |m\rangle =
(-1)^m(-i)^{n-1}{{[m+n]_q!} \over {[m]_q! \, [n]_q!}} |m+n\rangle \quad .        
\label{(A.3.5)} 
\ee 
so that as stated in our comments after \eq{(74)}, $z$ is real for odd $n$
and imaginary for even $n$. 

Turning next to the important 
result in the second line of \eq{(82)}, we calculate first the effect of the 
left side of this upon $|m\rangle \, , \, m=rn+p \,$, finding
$(r+1)|(r+1)n+p-1\rangle \, - \, z|rn+p-1\rangle $. If $p=0$, 
$|rn-1\rangle=|(r-1)n+(n-1)\rangle $, 
and computing the action of $z$ on it we get 
$((r+1)-r)|rn+n-1\rangle =|rn+n-1\rangle$.
If $p \neq 0$, there is no such shift down of the effective $r$-value, and,
applying \eq{(A.2)}, we
get zero, so that
\be
[{{\cal D}}_L \, , \, z] |rn+p\rangle = \delta_{p0}|(r+1)n+(p-1)\rangle 
=\delta_{p0}|rn +(n-1)\rangle
\quad . \label{(A.4)} 
\ee 
\noindent The right side of the identity in question is
\be
{{\theta^{n-1}} \over {[n-1]_q!}}|m\rangle = {{[m+n-1]_q!} \over {[m]_q!
\, [n-1]_q!}} |m+n-1\rangle \quad . \label{(A.5)} 
\ee 
\noindent  If $p=0$ the factor on the right of \eq{(A.5)} is $1$ and the right side 
of \eq{(A.5)} is $|(r+1)n-1\rangle=|rn+(n-1)\rangle$. 
If $p \neq 0$, the numerator of the factor on 
the right side of \eq{(A.5)} contains the factor $[(r+1)n]_q$, which vanishes. 
Since there is no factor $[n]_q$ in the denominator, the right side of \eq{(A.5)} is
then zero, and proof of the required identity is complete. 
One meets proofs along 
the same lines in various related contexts. We may also use \eq{(85)} to obtain 
the action of $\partial_\theta$ on ${{\cal V}}$ in the form
\be
\partial_\theta|rn+p\rangle =(1-\delta_{p0})|rn+(p-1)\rangle \quad . 
\label{(A.6)} 
\ee 

The above results hint at some sort of direct product structure emerging from
${{\cal V}}$ in the limiting case. To bring this fully into evidence, we write
\be
|rn+p\rangle \equiv |r,p\rangle \equiv |r\rangle \otimes |p\rangle \in
{{\cal V}}_{HO} 
\otimes {{\cal V}}_n \quad . \label{(A.7)} 
\ee 
Then the actions in ${{\cal V}}$ of $\theta ,z$ etc. can be  
presented in the form
\be
\theta \equiv 1 \otimes \theta_c \, , \, z \equiv z_c \otimes 1 \quad ,
\label{(A.8)} 
\ee 
\be
\partial_\theta \equiv 1 \otimes \partial_{\theta c} \, , \,
\partial_z \equiv \partial_{zc} \otimes 1 \quad , \label{(A.9)} 
\ee 
\be
{{\cal D}}_L \equiv 1 \otimes \partial_{\theta c} + \partial_{zc} \otimes 
{{\theta_c^{n-1}} \over {[n-1]_q!}} \quad , \label{(A.10)} 
\ee 
where 
\be
 z_c|r\rangle =(r+1)|(r+1)\rangle \,
, \, \partial_{zc}|r\rangle =|r-1\rangle \quad , \label{(A.11)} 
\ee 
\be
\theta_c|p\rangle =[p+1]|p+1\rangle \, , \, \partial_{\theta c}|p\rangle
=|p-1\rangle \quad . \label{(A.12)} 
\ee 
Thus $z_c$ and $\partial_{zc}$ have standard (Bargmann type) actions in 
${{\cal V}}_{HO}$, and $\theta_c$ and $\partial_{\theta c}$ are represented
by $n \times n$ matrices in ${{\cal V}}_n$. If in this appendix we had, as in \eq{(A.3.5)}, 
employed the choice 
\eq{(11.3)} instead of \eq{(6.7)} then results similar in nature but
different in some detail would have emerged. On the one hand $\theta_c$ would 
have been represented at stage \eq{(A.12)} by a real matrix, but, on the other, a 
variety of somewhat unattractive ($q$-dependent) phases would creep into 
certain results.

\section{}

In this appendix we give an example of how generalized Grassmann variables 
can be used to derive identities in $q$-analysis
\r{CP,GRA}. In general, 
\be 
(\epsilon+\theta)^r=\sr \epsilon^m\theta^{r-m}{[r]_q!\over[m]_q![r-m]_q!}\quad,
\label{(158)}
\ee
where $[\theta,\epsilon]_q=0$. To get $q$-binomial identities out of this we 
begin by replacing $\epsilon$ with $-\alpha q^{-N}\theta$, where $\alpha$ is 
an ordinary number). This clearly commutes with $\theta$ in the same way as 
$\epsilon$, so that from \eq{(158)} we have,
\be 
(\theta-\alpha q^{-N}\theta)^r=\sr (-\alpha
q^{-N}\theta)^m\theta^{r-m}{[r]_q!\over[m]_q![r-m]_q!}\quad.
\label{(159)}
\ee
Acting with both sides on the ground state $\zz$ of some representation with 
${\cal D}_L\zz=0$ we find that
\be 
\eqalign{
	\pr(1-\alpha q^{-m})
	&=\sr (-\alpha)^m q^{{1\over2}m(m-1)}q^{-mr}{[r]_q!\over[m]_q![r-m]_q!}
	\cr
	&=\sr (-\alpha)^m q^{-mr}{[r]_q!\over[m]_{q^{-1}}![r-m]_q!}\quad.}
\label{(160)}
\ee
When $q$ is a primitive $n$-th root of unity, and $r=n-1$, there is an 
interesting special case of this result. 
To derive it note first that in this case
\be 
[m]_q=-q^m[n-m]_q\quad,
\label{(161)}
\ee 
so that 
\be 
[m]_q!=(-1)^m q^{{1\over2}m(m+1)}{[n-1]_q!\over[n-m-1]_q!}\quad.
\label{(162)}
\ee
Then we have from \eq{(160)}
\be 
\eqalign{
	\pn(1-\alpha q^{-m})&=\sn(-\alpha)^m q^{{1\over2}m(m+1)} 
	{[n-1]_q!\over[m]_q![n-m-1]_q!}\cr
	&=\sn\alpha^m=[n]_{\alpha}\quad.\cr}
\label{(163)}
\ee
In particular, for $\alpha=1$ we obtain 
\be 
\pn(1- q^{-m})=n\quad,
\label{(164)}
\ee
which verifies the identity quoted after \eq{(125)}, 
and for $\alpha=-1$,
\be  
\pn(1+ q^{-m})={1-(-1)^n\over2}=\left\{\eqalign {& 0\hskip5pt\hbox{  
for\hskip5pt even\hskip5pt
 n,}\cr
& 1\hskip5pt\hbox{for\hskip5pt odd\hskip5pt n.}\cr}\right.
\label{(165)}
\ee

\section{}

In this appendix we discuss the Fock space representations of the $q$-deformed 
oscillators \r{AC,Macfarlane,Biedenharn} 
used in sections VI and XI. Let us first restate their 
definition,
\be 
[\am,\ap]_{q^{\pm {1\over2}}}=q^{\mp{N\over2}}\quad.
\label{(166)}
\ee
Our aim is to determine for which $q$ we can build Fock space representations in which
$a_{+}$ is the adjoint of $a_{-}$ in a space of positive definite metric:
\be  
(\ap)^\dagger=\am\quad,
\label{(167)}
\ee
in other words,  we aim to identify all of the $q$-oscillators with physical 
Fock space representations. From \eq{(166)} we have (see \eq{(6.6)})
\be 
\ap \am=[[N]]_q\quad,\quad\am\ap=[[N+1]]_q\quad,
\label{(168)}
\ee
so that of $\am$ and $\ap$ can be realized in terms of undeformed bosonic 
oscillators $b$ and $\bd$ satisfying $[b,\bd]=1$ as follows,
\be 
\ap=\left({[[N]]_q\over N}\right)^\alpha\bd=\bd
\left({[[N+1]]_q\over N+1}\right)^\alpha
\quad,\quad 
\am=b\left({[[N]]_q\over N}\right)^{1-\alpha}\quad.
\label{(170)}
\ee
The parts contained within parentheses are known as deformation
functions\r{CurtZa,Poly,KulDam,Song}. 
Matrix expressions containing the same information could be given, 
but this concise deformation function notation is more convenient to work 
with. On a Fock space with ground state $\vert 0\rangle$, such that 
$a_-\vert 0\rangle=0$ 
(which is equivalent to $b\vert 0\rangle=0$),
$N$ is hermitian 
and has nonnegative integer eigenvalues $m$.
If \eq{(167)} is to be satisfied then clearly we need
$\alpha={1\over2}$, in which case the requirement is that $[[N]]_q$ should be  
real and nonnegative   for each $m$. Since  $N^\dagger=N$, the reality 
condition gives $\qst=\qpm$, where $*$ denotes complex conjugation. Thus we 
have two cases to consider, $\q$ real and $\vert \q\vert=1$. We treat
$\q=-1$ as a third case since it involves the taking of a limit. 

If $\q$ is
real and negative, then it is clear that $[[N]]_q$ takes on alternately positive 
and negative values as $m$ increases violating the above condition. For $\q$ 
real and positive we can write $\q=e^s$, where s is real, so that
\be 
[[N]]_q={\sh (sN)\over \sh s}\quad,\label{(171)}
\ee
which is nonnegative for all $m$ if $s\geq0$.
\par 
When $\vert q^{1\over2}\vert=1$, we can 
write $\q=e^{{{2\pi i}\over r}}$, where $r$ is real. Then we have,
\be 
[[N]]_q={\ns ({2\pi N\over r })\over \ns ({2\pi\over r})}\quad.\label{(172)}
\ee
Note that for $r=\pm{2\beta\over\gamma}$, where $\beta$
and $\gamma$ are positive integers with $\beta$ prime relative to $\gamma$ , 
we have $[[\beta]]_q=0$, 
so that starting from $\zz$, only states with $0\leq m<\beta$
can be reached using $\ap$ and $\am$ (so that on the Fock space 
$(a_+)^{\beta}=(a_-)^{\beta}=0)$. 
This is important because, in general
the numerator of \eq{(172)} changes sign when $m$ passes ${r\over2}$, violating the
nonnegativity condition on $[[N]]_q$. However, for $r$ of the form given above, 
we know that for $\beta\leq {r\over2}$, such states
cannot be reached. The only solution has $\gamma=1$, so that $r=2\beta$.
Thus only for $\q=e^{{\pm{\pi i}\over \beta}}$ do the representations with $\vert\q\vert=1$ 
remain physical. 

The $\q\to1$ limit of $[[N]]_q$ 
is just the undeformed operator $N$, and from
this we obtain the $\q\to-1$ limit as follows
\be 
\eqalign{
	\qml \left({q^\N-q^{-\N}\over \q-\qn}\right)
	&=(-1)^{N-1}\qpl \left({q^\N-q^{-\N}\over \q-\qn}\right)\cr
	&=(-1)^{N-1}N\quad,\cr}
\label{(173)}
\ee
so that $\q=-1$ violates the nonnegativity condition, and must be excluded. 
To summarize, $[[N]]_q$ is real and nonnegative and hence 
\eq{(167)} is satisfied for a) $\q$ real and $\q\geq1$, and b) for $\q=e^{{\pm{\pi i}\over \beta}}$,
 where $\beta$ is a positive integer. When $\q$ takes on one of these allowed
values we can, from \eq{(167)}, write $\ap=\a$ and $\am=a$,
so that \eq{(166)} becomes 
\be 
[a,\ad]_{q^{\pm {1\over2}}}=q^{\mp{N\over2}}\quad,
\label{(174)}
\ee
in which $a$ and $\a$ have the implied hermiticty properties.
\\




\end{document}